# NP in BQP with Nonlinearity

Phil Gossett
Silicon Graphics, Inc. 2011 N. Shoreline Blvd. Mountain View, CA 94043-1389
e-mail:pg@engr.sgi.com
April 27, 1998

**Abstract:** If one modifies the laws of Quantum Mechanics to allow nonlinear evolution of quantum states, this paper shows that NP-complete problems would be efficiently solvable in polynomial time with bounded probability (NP in BQP). With that (admittedly very unlikely) assumption, this is demonstrated by describing a polynomially large network of quantum gates that solves the 3SAT problem with bounded probability in polynomial time. As in a previous paper by Abrams and Lloyd (but by a somewhat simpler argument), allowing nonlinearity in the laws of Quantum Mechanics would prove the "weak Church-Turing thesis" to be false. General Relativity is suggested as a possible mechanism to supply the necessary nonlinearity.

1. Introduction

Despite a series of successes for isolated problems (the Deutsch-Jozsa algorithm [1], Simon's problem [2], discrete log and factorization [3]), a number of recent papers [4,5,6,7,8,9] have cast doubt that quantum computers are capable of solving NP-complete problems in polynomial time (with either certainty or bounded probability). However, these papers have all fallen short of strict no-go proofs, except for the special case of so-called "black box" algorithms.

There are a number of reasons to suspect that NP-complete problems are, in fact, efficiently solvable by quantum computation with bounded probability in polynomial time. Maymin [10] has shown that the lambda-q calculus, modelled on quantum computation, can efficiently solve NP-complete problems in polynomial time. Very recently, Abrams and Lloyd [11] have shown that the introduction of nonlinearity into Quantum Mechanics would allow quantum computers to solve NP-complete problems in polynomial time, though by a somewhat more complicated argument than the one presented in this paper. And simulation of quantum systems by quantum computers has been shown to be efficient [12].

While this paper cannot be considered a proof, since it requires a rather unlikely modification to the laws of Quantum Mechanics to allow for nonlinear evolution of quantum states, this paper conditionally demonstrates that allowing nonlinearity would provide a proof that the 3SAT problem, one of the first examples of an NP-complete problem [13], is solvable by a quantum computer with bounded probability in polynomial time

## 2. Limits of Quantum Computation with Linearity

The quantum algorithms described in the literature typically have the following form: Start with some number of qubits in known pure states. Some or all of them are Hadamard transformed into a superposition of states. A series of unitary transforms are applied. Some, all, or none of the resulting qubits are Hadamard transformed to cause different possible computational paths to interfere. The resulting qubits are measured [14].

With the (quite reasonable) assumption of linearity, it has been shown that no "black box" quantum algorithm can gain more than a quadratic factor over classical algorithms [4,5,6]. Grover's search algorithm [15] is an example of this category of quantum algorithms. This quadratic limitation exists because, while there can be exponentially many superpositions of quantum states, the probability of success is only quadratically improved over a classical probabilistic Turing machine. As has been shown before [11], and is demonstrated below by a very simple construction, this limitation can be overcome if the strict linearity of Quantum Mechanics is abandoned.

(Note that it is not even strictly necessary - whether it's possible or not - for a useful quantum computer to itself be able to emulate a Turing machine. We already have classical computation universal machines. Appending a quantum computer onto a classical Turing machine doesn't make the Turing machine any less computation universal.)

## 3. Perfect One-hit Oracles

If we relax the requirement of quantum linearity, and allow the quantum state to evolve nonlinearly, it becomes possible to make a perfect one-hit oracle. A given superposition of states can be made to destructively interfere for all the wrong answers, and constructively interfere for the one right answer, producing the correct result with probability 1 (assuming ideal quantum devices). Given a perfect one-hit oracle, we're a short distance from a solution to the 3SAT problem, as is shown below.

To implement a perfect N qubit one-hit oracle, we first need a network of quantum gates [16] to implement an "inverse oracle function". This function answers "false" when the single correct N qubit input is supplied, and "true" when any incorrect input is supplied. All qubits of the inverse oracle function must, however, be ultimately left in an eigenstate, so as not to spoil the final measurement. First, a superposition of all possible oracle input states, together with a maximally entangled copy, is prepared in advance. Note that this preparation does not violate the "no-cloning" theorem [17], since the entangled pairs of qubits can be created maximally entangled [18]. An additional unentangled set of N qubits (uncorrelated with either copy of the oracle inputs) in a superposition of all possible states is also prepared in advance.

For each of N gates in succession (one per oracle input qubit), the inverse oracle function output is provided as one input of a controlled phase inversion gate. The corresponding unentangled qubit (in a perfect superposition of "true" and "false") is provided as the other input. The maximally entangled copy of the corresponding oracle input qubit is passed through. Looking at all three qubits together, when the two controlled phase inversion inputs are both "true", the phase of the entire 3 qubit state is inverted. The (obviously unitary) matrix corresponding to this is as follows:

$$\begin{pmatrix} 1 & 0 & 0 & 0 & 0 & 0 & 0 & 0 \\ 0 & 1 & 0 & 0 & 0 & 0 & 0 & 0 \\ 0 & 0 & 1 & 0 & 0 & 0 & 0 & 0 \\ 0 & 0 & 0 & 1 & 0 & 0 & 0 & 0 \\ 0 & 0 & 0 & 0 & 1 & 0 & 0 & 0 \\ 0 & 0 & 0 & 0 & 0 & 1 & 0 & 0 \\ 0 & 0 & 0 & 0 & 0 & 0 & -1 & 0 \\ 0 & 0 & 0 & 0 & 0 & 0 & 0 & -1 \end{pmatrix}$$

Note that while the phase produced by the controlled phase inversion gate will be "kicked back" into the oracle output qubit, a simple phase inversion does not change the logical ("true"/"false") value of the oracle output qubit, and hence does not disrupt its value for subsequent gates. Similarly, any phase kicked back into the initially unentangled qubits is irrelevant, since this will not disturb the eigenstates once these initially unentangled qubits are nonlinearly driven to |0>.

For each oracle input qubit in the correct state, the corresponding controlled phase inversion gate will not invert the phase, regardless of the state of the corresponding unentangled qubit. For each incorrect state, the phase will be inverted when the corresponding unentangled qubit is "true", and not when it is "false".

The (trivial) case of a one bit oracle is particularly easy to analyze. The input of the controlled phase inversion gate can be represented in "ket" notation as |fui>, where |f> is the inverse oracle function output qubit, |u> is the unentangled qubit, and |i> is one of the oracle input qubits. In this simple case, there are two possibilities:

1) If the inverse oracle function is "false" (|0>) when the input qubit is "false" (|0>), then |f> = |i>, and the state is:

(1/2) * (|000> + |010> + |101> + |111>).

Applying the unitary operator for the controlled phase inversion gate to this state produces:

(1/2) * (|000> + |010> + |101> - |111>).

If we now allow a nonlinear evolution of the state, which after (potentially many) applications of the nonlinearity, drives the |u> qubit to an eigenstate (for example, |0>) without disturbing the other qubits, the resulting state is:

(1/2) * (2*|000> + 0*|101>).

So output qubit |i> = |0> with probability 1.

2) If the inverse oracle function is "false" (|0>) when the input qubit is "true" (|1>), then |f> = not(|i>), and the state is:

(1/2) * (|001> + |011> + |100> + |110>).

Applying the unitary operator for the controlled phase inversion gate to this state produces:

(1/2) * (|001> + |011> + |100> - |110>).

If we now allow a nonlinear evolution of the state, which after (potentially many) applications of the nonlinearity, drives the |u> qubit to an eigenstate (for example, |0>) without disturbing the other qubits, the resulting state is:

(1/2) * (2*|001> + 0*|100>).

So output qubit |i> = |1> with probability 1.

In either case, the incorrect result cancels out perfectly, and the correct result reinforces perfectly, producing the correct result with probability 1. In this (trivial) case, there is no need for further transforms or nonlinearities, since all qubits are already in eigenstates.

For (non-trivial) cases with multi-bit oracles, the input of the multi-bit oracle can be represented in "ket" notation as |e1;e2;u> (reordered compared to the above trivial case), where |e1> and |e2> are the two sets of N pair-wise maximally entangled qubits, and the single instance of |u> is the set of N unentangled qubits, and the semicolons represent concatenation. The |e1> qubits are used for the inverse oracle function, with one of them being temporarily used as the output of the inverse oracle function, before being inverse transformed back to their original state. The |e2> qubits are conditionally phase inverted, and are ultimately measured. The state of these pair-wise maximally entangled plus unentangled inputs is:

```
              N-1  N-1
(1/2^N)  *    Σ    Σ   |e;e;u>.
              e=0  u=0
```

Let's further define the N bit correct answer for the one-hit oracle to be |c>. Then the 2^(3*N) by 2^(3*N) unitary matrix representing the overall system (where i and j each span the concatenated qubits |e1;e2;u>, and parity(|u>) is the N-way xor of the qubits of |u>) is:

```
U(i, j) =   1(i == j, |e1> == |c>, for 2^N values of |u>)
            1(i == j, |e1> != |c>, parity(|u>) == |0>)
           -1(i == j, |e1> != |c>, parity(|u>) == |1>)
            0(i != j).
```

If we now apply the operator U to the above pair-wise maximally entangled plus unentangled |e1;e2;u> state, and again nonlinearly drive the |u> qubits to eigenstates, with the |e1> qubits still maximally entangled with the |e2> qubits, we get output qubits |e2> = |c> with probability 1.

Again, the incorrect results all cancel out perfectly, and the one correct result reinforces perfectly, producing the correct result with probability 1.

4. Application of Nonlinearity

Rather than selecting a specific model of quantum nonlinearity, we will instead describe the properties it must have for this construction to work. For our purposes, such a quantum nonlinearity must be deterministic, be a function only of the orientation of a state on the Riemann sphere (ignoring arbitrary phase factors), and have non-negative observable probabilities with unit total probability ("reality"). Note that this last property is already guaranteed, since the above one-hit oracle construction yields probability 1 with the given input state, assuming the nonlinear driving of the |u> qubits to |0>.

The requirement that the nonlinearity be only a function of orientation on the Riemann sphere has two consequences. First, since the linear operator U used above is an identity to within a sign negation, the orientation on the Riemann sphere of each qubit is unaffected by this linear operator. So for the purposes of application of the nonlinearity, the qubits can be treated independently. Furthermore, since states with opposite sign are treated identically by the nonlinearity, anything that would have cancelled out before the nonlinearity will still cancel out after its application.

So the application of the nonlinearity is now straightforward. For each of the |u> qubits independently, rotate the orientation of the basis states (|0> and |1>) on the Riemann sphere by 90 degrees minus epsilon. This will leave the |0> state slightly "above" the "equator" of the Riemann sphere, and the |1> state slightly "below". (Note that the "equator" can be any geodesic on the Riemann sphere, picked as needed to maximize the nonlinearity.)

Assuming the nonlinearity is "smooth", the magnitude of its effect will be (approximately) linear with the magnitude of epsilon. (Note that any discontinuities only make things better, since the discontinuity can be made to straddle the equator by rotation.) So, assuming the nonlinearity is relatively small, applying the nonlinear operator will cause a rotation of the orientation of the |0> state relative to the |1> state, which applied a number of times proportional to the accuracy desired, will line up the |0> and |1> states to be approximately 2*epsilon apart, across the equator of the Riemann sphere. The resulting approximately

identical states are then rotated back to the |0> basis. We have thus nonlinearly driven the |u> qubits to the |0> state in a time linearly proportional to the accuracy desired.

Note that there is a possible, though impractical, physical implementation of this nonlinearity. In the presence of an appropriately oriented magnetic field, the energy of the |1> basis state would be higher than the energy of the |0> basis state. One could argue that this energy difference would cause a (tiny) change in the mass-energy of the qubit. This in turn would cause an incredibly small time dilation of the evolution of the qubit, providing the nonlinearity by General Relativity. But this would be such a tiny effect, that the quantum state would almost certainly have decohered long before the nonlinearity had any significant effect.

5. Extension to 3SAT

Paralleling the construction of the one-hit oracle described above, we construct an oracle (not necessarily one-hit) with one more bit than the total number of inputs to the boolean equation to be tested for satisfiability. If this extra qubit is |0>, we produce the logical negation of the boolean equation to be tested as the inverse oracle function output. If the extra qubit is |1>, we output |0> if all the other oracle input qubits are also |1>, and output |1> if some or all of the other oracle input qubits are |0>. Note that since the boolean equation to be tested in the 3SAT problem can only be polynomial in size [13], this network can only be polynomial in size, and therefore in depth, and hence (assuming constant delay per gate) can only take polynomial time to evaluate.

If the boolean equation under test is not satisfiable, the inverse oracle function will only be |0> when all the oracle inputs (including the extra one) are |1>. If this is the case, the network of quantum gates described above will always produce all output qubits as |1>. If there is one set of input values that satisfies the boolean equation under test, then half of the time the extra qubit will be |1> when measured, and half the time it will be |0>. If there are more than one sets of input values that satisfy the boolean equation under test, then the extra qubit will be |0> with a correspondingly larger probability.

By repeating the computation M times, if the extra qubit is always |1>, then the boolean equation under test is not satisfiable with probability $1 - 2^{-M}$. If the extra qubit is ever |0>, the boolean equation under test is satisfiable with probability 1.

(Note that while this is a probabilistic algorithm, the probability can be made arbitrarily close to unity (at the cost of a modest constant factor of M). The probability of error can be made smaller than the error rate of whatever physical devices this quantum computer is made out of. So, for all practical purposes, this probabilistic algorithm is as good as deterministic.)

Thus, we have demonstrated, under the assumption of quantum nonlinearity, solution of the NP-complete 3SAT problem with (arbitrarily) bounded probability in polynomial space and time.

6. Conclusions

Landauer [19] has made the observation that computability is consequence of the laws of Physics rather than of Mathematics. As Deutsch [20] has pointed out, if Physics did not permit ordinary arithmetic operations, these would be considered "noncomputable" functions, and proofs that relied on them would be considered "nonconstructive" proofs. It should therefore not be too surprising that the so-called "weak Church-Turing thesis" (also known as the "Cook-Karp thesis" [21]) should fail when computation is considered under quantum, rather than classical, mechanics.

The construction described in this paper demonstrates the solution in polynomial space and time (with bounded probability) of a well-known NP-complete problem (3SAT) by a quantum computer, under the assumption of quantum nonlinearity. Since (despite everyday appearances) we live in a quantum rather classical universe [22], nonlinearity in Quantum Mechanics could be exploited to refute the weak Church-Turing thesis.

Unfortunately for this conditional disproof of the weak Church-Turing thesis, experimental evidence [23] demonstrates the linearity of Quantum Mechanics to remarkable precision. Still, the success of quantum factorization, and the equally remarkable failure of attempts to reduce classical factorization to polynomial (probabilistic or deterministic) complexity leave hope for some future refutation of the weak Church-Turing thesis, perhaps using some "classical" (General Relativistic or otherwise) nonlinear effect to somehow evade the strictly linear "black-box" construction.